\documentclass[pre,twocolumn,aps]{revtex4}

\usepackage{graphicx}
\usepackage{bm}
\usepackage{amsmath}
\begin{document}
\title{Shifted forces in molecular dynamics}
\author{S{\o}ren Toxvaerd and Jeppe C. Dyre}
\email{dyre@ruc.dk}
\affiliation{DNRF Centre ``Glass and Time'', IMFUFA, Department of Sciences, Roskilde University, Postbox 260, DK-4000 Roskilde, Denmark}
\date{\today}

\begin{abstract}
Simulations involving the Lennard-Jones potential usually employ a cut-off at $r=2.5\sigma$. This paper investigates the possibility of reducing the cut-off. Two different cut-off implementations are compared, the standard shifted potential cut-off and the less commonly used shifted forces cut-off. The first has correct forces below the cut-off, whereas the shifted forces cut-off modifies Newton's equations at all distances. The latter is nevertheless superior; we find that for most purposes realistic simulations may be obtained using a shifted-forces cut-off at $r=1.5\sigma$, even though the pair force is here 30 times larger than at $r=2.5\sigma$. 
\end{abstract}

\maketitle

Molecular dynamics (MD) simulations solve Newton's equations of motion by discretizing the time coordinate. The time-consuming part of any MD simulation is the force calculation. For a system of $N$ particles this is an $O(N^2)$ process whenever all particles interact. In practice the interactions are negligible at long distances, however, and for this reason one always introduces a cut-off at some interparticle distance $r=r_c$ beyond which interactions are ignored \cite{sim}. 

The standard Lennard-Jones (LJ) pair potential is given by

\begin{equation}\label{LJ}
u_{\rm LJ}(r)
\,=\,4\varepsilon  \left[(\sigma/r)^{12}-(\sigma/r)^{6}\right]\,.
\end{equation}
Usually, a cut-off at $r_c=2.5\sigma$ is employed; at this point the potential is merely 1.6\% of its value at the minimum ($-\varepsilon$). Although a cut-off makes the force calculation an almost $O(N)$ process, this calculation remains the most demanding in terms of computer time. 

The present paper investigates the possibility of reducing the LJ cut-off below $2.5\sigma$ without compromising accuracy to any significant extent. Before presenting evidence that this is possible, it is important to recall that quantities depending explicitly on the free energy are generally quite sensitive to how large is the cut-off. Examples include the location of the critical point \cite{smi92}, the surface tension \cite{smi92,gro09}, and the solid-liquid coexistence line \cite{val07,ahm10}. For such quantities even a cut-off at $2.5\sigma$ gives inaccurate results, and in some cases the cut-off must be larger than $6\sigma$ to get reliable results \cite{gro09}. Note, however, that if a simulation gives virtually correct particle distribution, the thermodynamics of for instance coexisting phases can be accurately calculated by application of standard first-order perturbation theory \cite{fop}.  

This note relates to systems for which the standard cut-off at $2.5\sigma$ gives a satisfactory radial distribution function. We compared two cut-off implementations at varying cut-off's with the ``true'' LJ system, the latter being defined here by the cut-off $r_c=4.5\sigma$. One cut-off is the standard ``truncated and shifted potential'' (SP for shifted potential), for which the radial force is given \cite{sim} by ($f_{\rm LJ}(r)=-u'_{\rm LJ}(r)$ is the LJ radial force)

\begin{equation}\label{shp}
f_{\rm SP}(r)\,=\,
\begin{cases}
f_{\rm LJ}(r) & \text{if}\,\,  r<r_c \\
0 &  \text{if}\,\, r>r_c\,\,.
\end{cases}
\end{equation}
This is referred to as a SP cut-off because it corresponds to shifting the potential below the cut-off and putting it to zero above, which ensures continuity of the potential at $r_c$ and avoids an infinite force here.

The ``truncated and shifted forces'' cut-off (SF for shifted forces) \cite{sim,sf} has the force go continuously to zero at $r_c$, which is obtained by subtracting a constant term:

\begin{equation}\label{shf}
f_{\rm SF}(r)\,=\,
\begin{cases}
f_{\rm LJ}(r) - f_{\rm LJ}(r_c) & \text{if}\,\,  r<r_c \\
0 &  \text{if}\,\, r>r_c\,\,.
\end{cases}
\end{equation}
This corresponds to the following modification of the potential: $u_{\rm SF}(r)=u_{\rm LJ}(r) -(r-r_c) u'_{\rm LJ}(r_c)-u_{\rm LJ}(r_c)$ for $r<r_c$, $u_{\rm SF}(r)=0$ for $r>r_c$. Use of a SF cut-off has recently become popular in connection with improved methods for simulating systems with Coulomb interactions \cite{coulomb}.

\begin{figure}
  \centering
  \includegraphics[width=100mm,angle=-90]{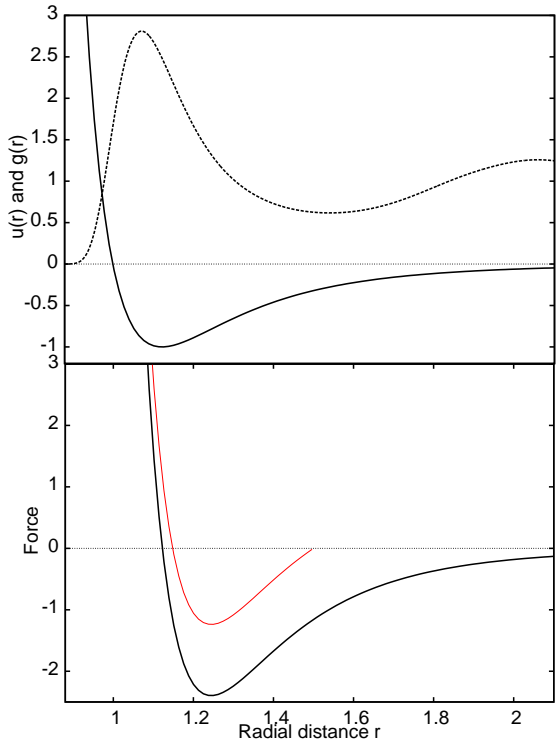}
 \caption{\label{basics}
(a) The Lennard-Jones potential (black full curve) and the radial distribution function $g(r)$ (black dashed curve) for a system at $\rho=0.85$ and $T$=1.00 in dimensionless units.
(b) The radial force, $f_{\rm LJ}(r)=-u_{\rm LJ}'(r)$ (black). At $r=1.5\sigma$ the force is 30 times larger than at $r=2.5\sigma$. Also shown is the shifted force (SF) for a cut-off at $1.5 \sigma$ (red).
}
\end{figure}

 \begin{figure}
  \centering
\includegraphics[width=60mm,angle=-90]{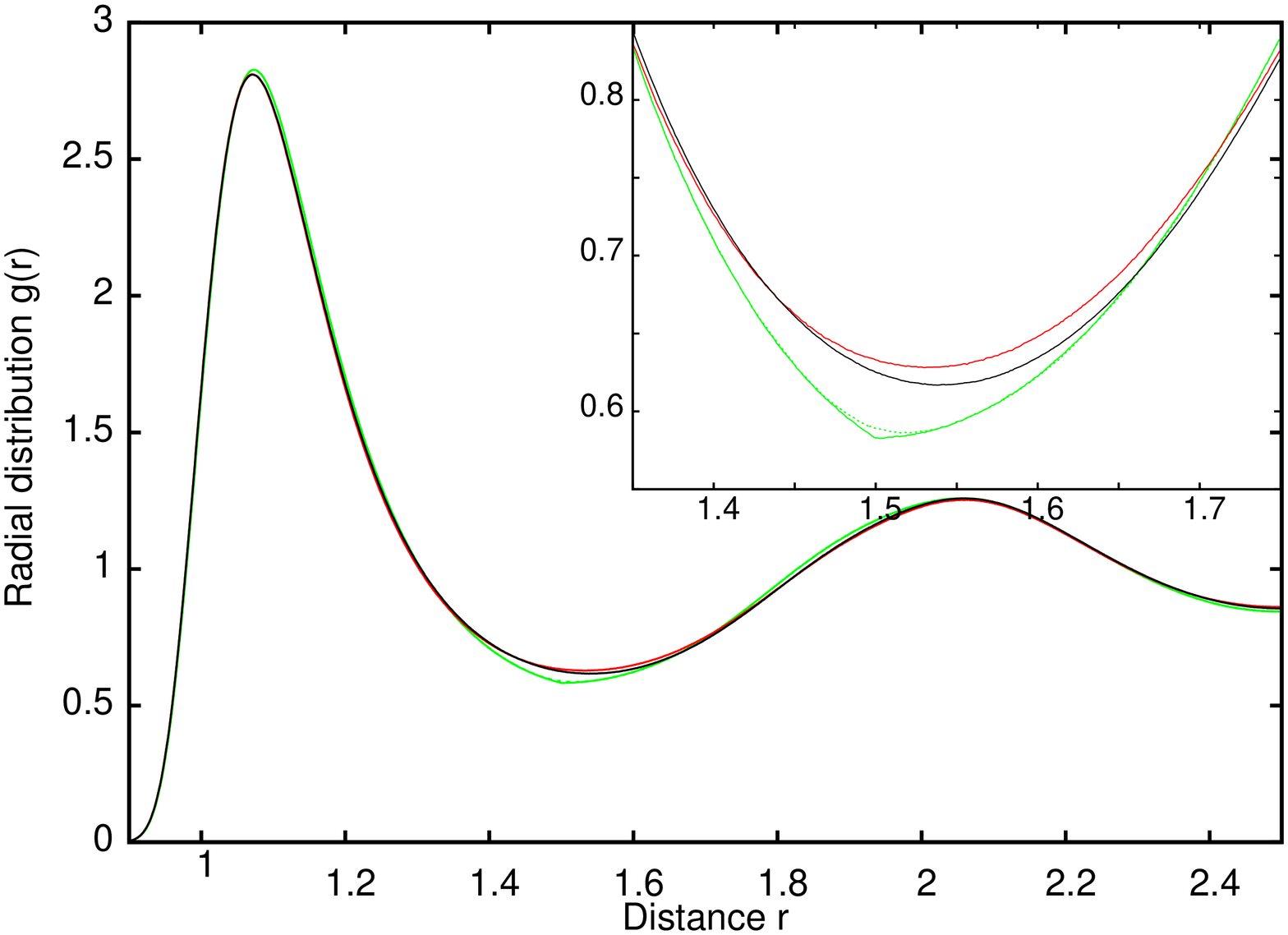}
 \caption{\label{gr_fig}
Radial distribution function $g(r)$ for the ``true'' LJ system (black) and two cut-off's at $r_c=1.5\sigma$. The red curve gives results for a SF cut-off, the green curve for a SP cut-off. The green dashed curve gives results for a SP cut-off with smoothing of the force and its derivative at the cut-off  \cite{smoothe}; this however does not improve the SP results.
}
\end{figure}

\begin{figure}
  \centering
  \includegraphics[width=60mm,angle=-90]{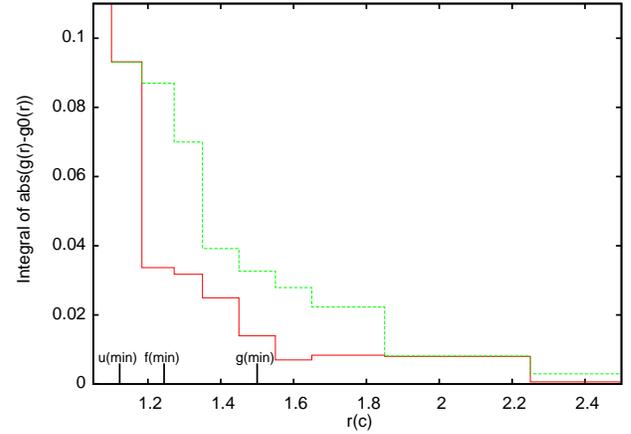}
 \caption{\label{hist_fig} Integrated numerical difference $\int_0^{4.5\sigma} |g(r)-g_0(r)| dr$ of the true radial distribution function, $g_0(r)$, and $g(r)$ for various cut-off distances $r_c$. The red curve gives results for the SF cut-off, the green for the SP cut-off. Smoothing a SP cut-off \cite{smoothe} does not improve its accuracy (data not shown).
}
\end{figure}

We simulated the standard single-component LJ liquid at the state point that in dimensionless units has density $\rho=0.85$ and temperature $T=1.0$ \cite{toxmd}. This is a typical moderate-pressure liquid state point \cite{sim,verlet}. Other state points were also examined -- including several state points of the fcc crystal, at the liquid-gas interface, at the solid-liquid interface, and for a supercooled system -- leading in all cases to the same overall conclusions. For this reason we report below results for just one state point of the LJ liquid and one of the Kob-Andersen binary LJ (KABLJ) liquid \cite{kablj}. 2000 LJ particles were simulated using the standard central-difference $NVT$ and $NVE$ algorithms (Figs. 2, 3 and 4, 6, respectively); 1000 particles of the KABLJ liquid were simulated using the $NVT$ algorithm (Fig. 5).

Figure \ref{basics} shows the basics of the LJ system. In the upper figure the black curve gives the LJ pair potential $u_{\rm LJ}(r)$ and the black dashed curve the radial distribution function $g(r)$, which has its maximum close to $u$'s minimum. In the lower figure the black curve shows the LJ pair force $f_{\rm LJ}(r)$. The red curve gives $f_{\rm SF}(r)$ when a cut-off at $1.5\sigma$ is introduced; note that the shifted force differs significantly from the true force. 

Figure \ref{gr_fig} shows the true pair-distribution function (black) and the simulated $g(r)$ for three $r_c=1.5 \sigma$ cut-off's: SF (red), SP (green), and a smoothed SP cut-off ensuring the force and its first derivative go continuously to zero at the cut-off \cite{smoothe} (green dashed curve). The curves deviate little, except near the cut-off where the smallest errors are found for a SF cut-off (inset).

\begin{figure}
  \centering
  \includegraphics[width=60mm,angle=-90]{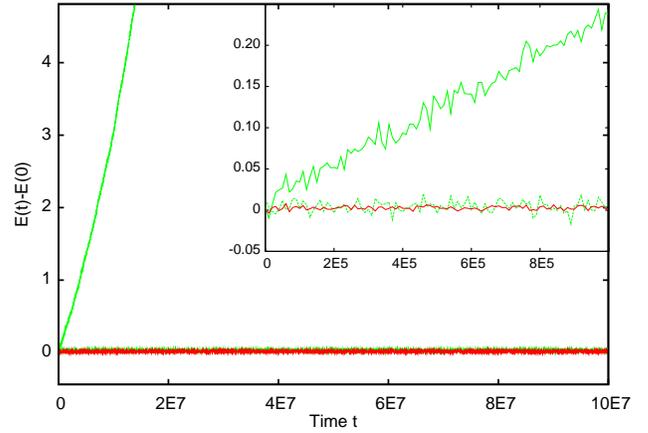}
 \caption{\label{en_fig}
Energy drift as a function of time for long simulations ($10^8$ time steps with $h=0.005$) with a cut-off at $1.5 \sigma$. The red curve gives results for the SF cut-off, the green for the SP cut-off, and the green dashed curve for a smoothed SP cut-off. Smoothing a SP cut-off stabilizes the algorithm, but the fluctuations are still somewhat larger than for a SF cut-off. The inset shows the initial part of the simulation.
}
\end{figure}

In order to systematically compare the SP and SF cut-off's we studied the LJ liquid for a range of cut-off's. Figure \ref{hist_fig} quantifies the difference between the computed $g(r)$ and the true, $g_0(r)$, by evaluating $\int_0^{4.5\sigma}|g(r)-g_0(r)|dr$. SF is red, SP is green. SF works better than SP for all values of $r_c$ above the ``WCA'' cut-off at the potential energy minimum \cite{fop} where SF=SP ($r_c= 2^{1/6} \sigma=1.12\sigma$). Smoothing a SP cut-off has only marginal effect compared to not smoothing it (results not shown). Applying first-order pertubation theory with the $g(r)$ obtained in a simulation with SF cut-off at $r_c=1.5\sigma$ leads to a pressure that deviates only 1\% from the correct value. 

Figure \ref{en_fig} studies energy drift in long $NVE$ simulations for $r_c=1.5\sigma$. The SF cut-off (red) exhibits no energy drift, whereas SP (green) does. Figure \ref{en_fig} also gives results when the force of a SP cut-off is smoothed \cite{smoothe} (green dashed curve). This leads to much better energy conservation \cite{sim}, but the energy fluctuations are somewhat larger  than for a SF cut-off. The simulations indicate the existence of a hidden invariance in the central-difference algorithm for a continuous force field deriving from a ``shadow Hamiltonian'' \cite{toxshadow}. 

\begin{figure}
  \centering
  \includegraphics[width=60mm,angle=-90]{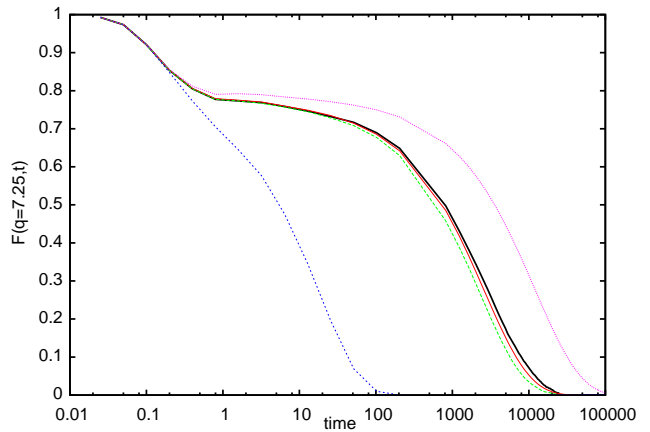}
 \caption{\label{skt_fig} 
The AA particle incoherent intermediate scattering function for the KABLJ liquid in the highly viscous regime ($T=0.45$, $\rho=1.2$, $q=7.25$). The figure shows results for the ``true'' system (black full curve), a SF cut-off at $1.5\sigma_{AA}$ (red full curve), a SP cut-off at $2.5\sigma_{AA}$ (green dashed curve), a SP cut-off at $1.5\sigma_{AA}$ (purple dotted curve), and for WCA cut-off (blue dashed curve).
}
 \end{figure}

Not only static quantities, but also the dynamics are affected little by replacing a $2.5\sigma$ SP cut-off with a $1.5\sigma$ SF cut-off. This is demonstrated in Fig. \ref{skt_fig}, which compares these two cut-off's for simulations of the incoherent intermediate scattering function of the supercooled KABLJ liquid \cite{kablj}. For reference a WCA cut-off simulation is included (blue dashed curve), which was recently shown to be inaccurate despite the fact that the WCA radial distribution function is reasonably good for this system \cite{wca_kablj}. A SP cut-off at $r_c=1.5\sigma_{AA}$ gives too slow dynamics (purple dotted curve). Within the numerical uncertainties incoherent scattering functions are identical for the ``true'' system, a SP cut-off at $r_c=2.5\sigma_{AA}$, and a SF cut-off at $r_c=1.5\sigma_{AA}$. Similar results were found for the single-component LJ liquid's dynamics. We conclude that a SF cut-off at $r_c=1.5\sigma$ generally works well for both statics and dynamics of LJ systems.

\begin{figure}
  \centering
  \includegraphics[width=60mm,angle=-90]{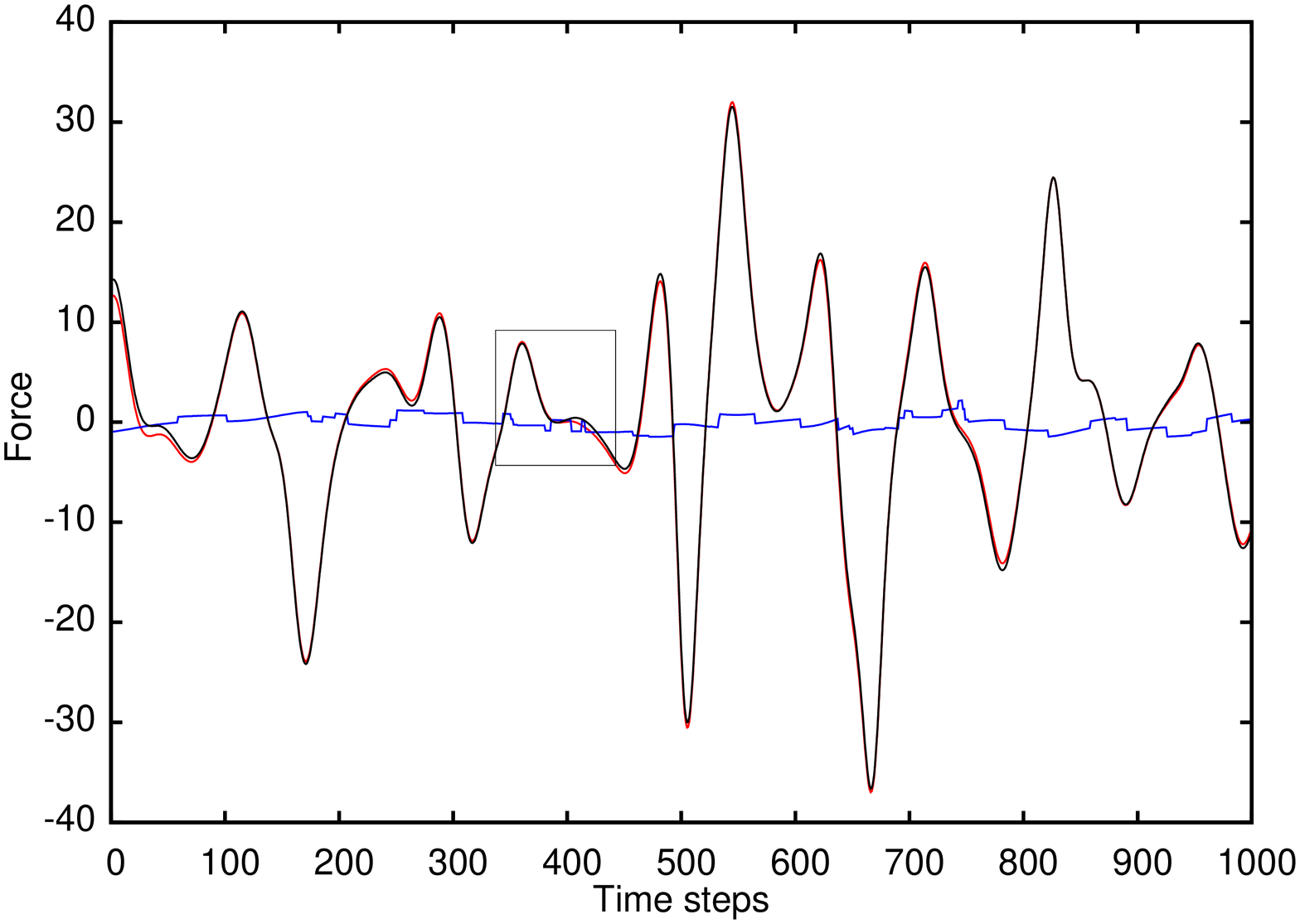}
  \includegraphics[width=60mm,angle=-90]{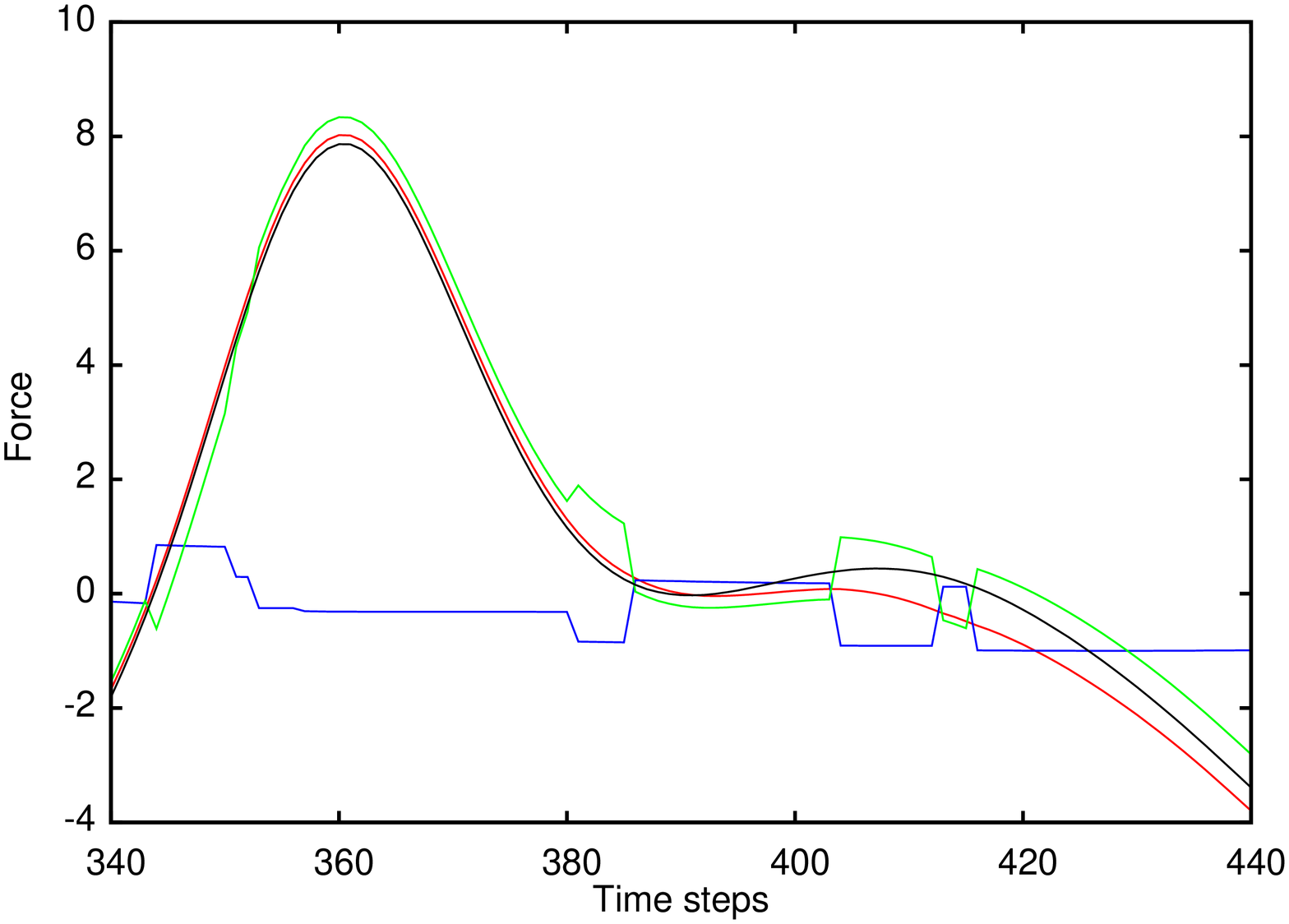}
 \caption{\label{ft_fig} 
(a) The x-component of the force on a typical particle during 1000 time steps. The black curve gives the true force, the red curve the force for a SF cut-off at $1.5 \sigma$. The blue curve gives the sum of the x-coordinates of the constant ``shift'' terms of Eq. (\ref{shf}). 
(b) Details after 340 steps. The green curve gives the SP force ($r_c=1.5 \sigma$). Only true and SF forces are smooth functions of time.
}
 \end{figure}

Why does a cut-off, for which forces are modified at all distances (SF), work better than when forces are correct below the cut-off (SP)? A SF cut-off modifies the pair force by adding a constant force for all distances below $r_c$; at the same time SF ensures that the pair force goes continuously to zero at $r=r_c$. Apparently, ensuring continuity of the force -- and thereby that $u''(r)$ does not spike artificially at the cut-off -- is more important than maintaining the correct pair force below the cut-off. How large is the change induced by the added constant force of the SF cut-off? Figure \ref{ft_fig} shows the x-component of the force on a typical particle as a function of time ($r_c=1.5 \sigma$). The black curve gives the true force, the red curve the SF force, and the blue curve the SF correction term. Although the true and SF individual pair forces differ significantly (Fig. \ref{basics}), the difference between true and SF total forces is small and stochastic ($\sim$3\%). This reflects an almost cancellation of the correction terms deriving from the fact that the nearest neighbors are more or less uniformly spread around the particle in question. It was recently discussed why adding a linear term ($\propto r$) to a pair potential hardly affects dynamics \cite{scl} and statistical mechanics  \cite{iso}: For a given particle's interactions with its neighbors the linear terms sum to almost a constant because, if the particle is moved, some nearest-neighbor distances increase and others decrease in such a way that their sum is almost constant. 

Figure \ref{ft_fig}(b) shows details of Fig. \ref{ft_fig}(a); we here added the SP force for the same cut-off (green). Both SP and correction terms are discontinuous; they jump whenever a particle pair distance passes the cut-off. Altogether, Fig. \ref{ft_fig} shows that not only does the sum of the constant forces on a given particle from its neighbors cancel to a high degree, so do the interactions with particles beyond the cut-off. The result is that the particle distribution is affected little by the long-range attractive forces, a fact that lies behind the success of perturbation theory \cite{fop,bh,tox71}.

In summary, when a SF cut-off is used instead of the standard SP cut-off, errors are significantly reduced. Our simulations suggest that a SF cut-off at $1.5\sigma$ may be used whenever the standard SP cut-off at $2.5\sigma$ gives reliable results; this applies even though the pair force at $r=1.5\sigma$ is 30 times larger than at $r=2.5\sigma$. A cut-off at $1.5\sigma$ is large enough to ensure that all interactions within the first coordination shell are taken into account (Fig. \ref{gr_fig}). Use of a $1.5\sigma$ SF cut-off instead of a SP cut-off at $2.5\sigma$ leads potentially to a factor of $(2.5/1.5)^3=4.7$  shorter simulation time for LJ systems.

\acknowledgments
The centre for viscous liquid dynamics ``Glass and Time'' is sponsored by the Danish National Research Foundation (DNRF).


\begin{thebibliography}{99}
\bibitem{smoothe} The radial force $f_{\rm SP}(r)$ was smoothed in the interval $r_c \pm \delta $ by replacing it with the function $f(r)=f(r_c -\delta)h(x)$ with $h(x)=1+ cx-(3+2c)x^2+(2+c)x^3$, where $x=(r-r_c +\delta)/2\delta$ and $c=2 \delta u''(r_c- \delta)/u'(r_c -\delta)$. This ensures that $f_{\rm SP}$ and $f'_{\rm SP}$ go smoothly to zero in the cut interval. In the simulations $\delta=0.025$ (other values lead to simular conclusions).
\end{thebibliography}
\end{document}